# Prospects for the discovery of the next new element: Influence of projectiles with $Z > 20$

C M Folden III[1], D A Mayorov[1,2], T A Werke[1,2], M C Alfonso[1,2], M E Bennett[1], and M J DeVanzo[1,3]

[1] Cyclotron Institute, Texas A&M University, College Station, TX 77843 USA
[2] Department of Chemistry, Texas A&M University, College Station, TX 77842 USA
[3] Department of Physics, Astronomy, and Geosciences, Towson University, Towson, MD 21252 USA

Email: Folden@comp.tamu.edu

**Abstract**. The possibility of forming new superheavy elements with projectiles having $Z > 20$ is discussed. Current research has focused on the fusion of $^{48}$Ca with actinides targets, but these reactions cannot be used for new element discoveries in the future due to a lack of available target material. The influence on reaction cross sections of projectiles with $Z > 20$ have been studied in so-called analog reactions, which utilize lanthanide targets carefully chosen to create compound nuclei with energetics similar to those found in superheavy element production. The reactions $^{48}$Ca, $^{45}$Sc, $^{50}$Ti, $^{54}$Cr + $^{159}$Tb, $^{162}$Dy have been studied at the Cyclotron Institute at Texas A&M University using the Momentum Achromat Recoil Spectrometer. The results of these experimental studies are discussed in terms of the influence of collective enhancements to level density for compound nuclei near closed shells, and the implications for the production of superheavy elements. We have observed no evidence to contradict theoretical predictions that the maximum cross section for the $^{249}$Cf($^{50}$Ti, $4n$)$^{295}$120 and $^{248}$Cm($^{54}$Cr, $4n$)$^{298}$120 reactions should be in the range of 10–100 fb.

## 1. Introduction

*1.1. Recent searches for new elements*
The last fifteen years have seen a tremendous growth in the periodic table, as six elements with $Z = $ 113–118 have been discovered by a collaboration of researchers from the Joint Institute for Nuclear Research (JINR) in Dubna, Russia and Lawrence Livermore National Laboratory (LLNL) in Livermore, California, USA. (See [1] for a review and [2-8] and references therein for details). These studies have used neutron-rich $^{48}$Ca projectiles reacting with targets of actinide nuclei. These data have shown that the *xn* cross sections for these reactions are roughly constant with increasing $Z$ of the product (see Fig. 7b in [1]), a result which is not currently understood.

There is significant interest in discovering even heavier elements, but such elements cannot be synthesized using $^{48}$Ca projectiles because sufficient quantities of the appropriate target elements are not available. Thus, the projectile atomic number $Z_p$ must be greater than 20, and a number of such experiments have been performed in recent years. For example, the JINR-LLNL collaboration has attempted to produce element 120 in the $^{244}$Pu($^{58}$Fe, $xn$)$^{302-x}$120 reaction, but observed no events with a one-event upper limit cross section of 0.4 pb [9]. At the GSI Helmholtzzentrum für Schwerionenforschung (GSI), an experiment has been conducted to search for element 120 in the

$^{248}$Cm($^{54}$Cr, xn)$^{302-x}$120 reaction, but also observed no events with an upper limit cross section of 0.56 pb [10]. The latter collaboration also plans to attempt the production of element 119 in the $^{248}$Cm($^{51}$V, xn)$^{299-x}$119 reaction [10]. A separate collaboration, also working at GSI, has attempted to produce elements 119–120 in the $^{249}$Bk($^{50}$Ti, xn)$^{299-x}$119 [11] and $^{249}$Cf($^{50}$Ti, yn)$^{299-y}$120 reactions, respectively. As of this writing, these authors are not aware of any reports of decay chains consistent with the production of superheavy elements (SHEs) in the $^{54}$Cr-, $^{51}$V-, or $^{50}$Ti-induced reactions.

*1.2. Analog reactions*

The "warm fusion" experiments described above require very long irradiations using the most sensitive equipment available because of the very small expected cross sections for production of evaporation residues (EVRs). With presently available technology, only a few atoms could possibly be observed during the irradiation, making a systematic study of these reactions difficult. These reactions typically produce a compound nucleus (CN) with an excitation energy $E^* \approx$ 35–45 MeV, followed by the emission of 3–5 neutrons. The projectile energy is typically several MeV above the Coulomb barrier, as opposed to so-called "cold fusion" reactions which require sub-barrier fusion. (See [12] for a discussion). The emission of neutrons occurs at a temperature of $\approx$1 MeV, and continues until the excitation energy of the residue falls below the neutron binding energy. During this de-excitation process, it is possible for the nucleus to fission, which is undesirable but highly probable. Therefore, a high fission barrier is likely to improve the survival probability, so increased shell stabilization is desirable. Recent element discoveries have produced CN near the next doubly magic shell, which is variously predicted to be found at $Z$ = 114 and $N$ = 184 [13, 14], $Z$ = 120 and $N$ = 172 [15], and $Z$ = 126 and $N$ = 184 [14, 16]. These closed shells are all predicted to result in nuclei with low ground-state deformations, which may result in an increased sensitivity to collective effects (see discussion below).

If one wishes to study the influence of projectiles with $Z_p$ > 20 using reactions with larger cross sections, it is necessary to ensure that several criteria are met. First, the reaction should occur above the barrier. Second, the total energy available for neutron emission should be consistent with that available in SHE synthesis. Third, the CN should be shell-stabilized and have some degree of similarity in deformation compared to SHEs. Fourth, the residues should have a large $\alpha$-decay branch to make them easy to detect in a silicon detector. At the Cyclotron Institute at Texas A&M University, we have studied so-called warm fusion "analog reactions." In these reactions, carefully chosen projectile and target nuclei react to form a CN which can mimic those produced in SHE studies. Table 1 shows a comparison of two reactions of interest to this study, along with corresponding data for similar reactions that have or could lead to the production of SHEs. All four reactions involve even-even nuclei reacting to produce a CN near a closed shell, and all four occur above the barrier. The center-of-mass energy of the projectile $E_{cm}$ was estimated by summing the neutron binding energies with an average neutron kinetic energy, assumed to be equal to twice the nuclear temperature at each de-excitation stage. Although the excitation energies for the $^{162}$Dy-target reactions are ≈7 MeV higher than for the $^{248}$Cm-target reactions, the total energy available for kinetic energy of the emitted neutrons (final column) is similar across all reactions.

**Table 1.** Comparison of "analog reactions" studied in this work with those expected to lead to the production of SHEs. $N_{CN}$ is the neutron number of the CN, $E_{cm}$ is the center-of-mass projectile energy at the measured or expected peak of the excitation function, $B_{Coul}$ is the Coulomb barrier calculated according to [12], $E^*_{CN}$ is the excitation energy of the CN, and $B_{n,i}$ represents the binding energy of each neutron [17] during the de-excitation process.

| Reaction | Ref. | $N_{CN}$ | $E_{cm} - B_{Coul}$ (MeV) | $E^*_{CN}$ (MeV) | $E^*_{CN} - \Sigma B_{n,i}$ (MeV) |
|---|---|---|---|---|---|
| $^{162}$Dy($^{48}$Ca, 4n)$^{206}$Rn | This work | 124 | 5.5 | 48 | 15.6 |
| $^{162}$Dy($^{54}$Cr, 4n)$^{212}$Th | This work | 126 | 10.2 | 50 | 15.8 |
| $^{248}$Cm($^{48}$Ca, 4n)$^{292}$Lv | [8] | 180 | 9.6 | 41 | 15.5 |
| $^{248}$Cm($^{54}$Cr, 4n)$^{298}$120 | [18] | 182 | 12.3 | 43 | 15.9 |

As a way to gain insight into the warm fusion reaction, we have performed a baseline excitation function measurement of the $^{162}$Dy($^{48}$Ca, $xn$)$^{210-x}$Rn ($x$ = 4,5) reactions, along with systematic studies of the excitation functions of the $^{159}$Tb($^{45}$Sc, $4n$)$^{200}$Rn, $^{162}$Dy($^{45}$Sc, $4n$)$^{203}$Fr, $^{159}$Tb($^{50}$Ti, $4n$)$^{205}$Fr, and $^{162}$Dy($^{54}$Cr, $4n$)$^{212}$Th reactions. Preliminary results and their implications for the production of SHEs are reported in the remainder of this article.

## 2. Experimental methods

The experiments were performed using the Momentum Achromat Recoil Spectrometer (MARS) [19, 20] located at the Texas A&M University Cyclotron Institute. The experimental setup was substantially similar to that described in [21], where the spectrometer efficiency for transmission of heavy element recoils was characterized. Enriched $^{48}$Ca [(91.0 ± 0.3)%] as $^{48}$CaCO$_3$ and $^{54}$Cr [(99.8 ± 0.1)%] as metal powders were purchased from Isoflex USA, San Francisco, California. Enriched $^{50}$Ti (≈55%) was purchased as a metal oxide from the same supplier and chemically reduced to a single metal piece. Monoisotopic $^{45}$Sc was used as a metal powder. In a series of temporally separated experiments, projectile beams of $^{48}$Ca$^{6+}$, $^{45}$Sc$^{6+}$, $^{50}$Ti$^{7+}$, and $^{54}$Cr$^{7+}$ were injected from the 6.4-GHz Electron Cyclotron Resonance source to the K500 cyclotron and accelerated to energies of 214.6, 224.6, 245.2, and 273.2 MeV, respectively. The energy of each beam was determined by passing it through a thin $^{nat}$C foil mounted downstream of the primary target position and measuring the magnetic rigidities of the resulting charge states in the first dipole (D1) of MARS. This procedure has an estimated uncertainty of ≈1%. Beam energies were varied using 0 (no degrader), 1.22, 2.25, 3.45, 4.50, 6.29, and 8.54 μm Al degraders located upstream of the target position. A 403-μg/cm$^2$ $^{162}$Dy target (with 75-μg/cm$^2$ $^{nat}$C upstream backing) was bombarded by $^{48}$Ca, $^{45}$Sc, and $^{54}$Cr beams to measure the excitation functions for the $^{162}$Dy($^{48}$Ca, $xn$)$^{210-x}$Rn, $^{162}$Dy($^{45}$Sc, $yn$)$^{207-y}$Fr and $^{162}$Dy($^{54}$Cr, $zn$)$^{216-z}$Th reactions, respectively. A 498-μg/cm$^2$ self-supporting $^{159}$Tb target was bombarded by $^{45}$Sc and $^{50}$Ti to measure the $^{159}$Tb($^{45}$Sc, $xn$)$^{204-x}$Rn and $^{159}$Tb($^{50}$Ti, $yn$)$^{209-y}$Fr excitation functions, respectively. The beam dose was monitored by Rutherford scattering using two circularly collimated Si detectors mounted equatorially at ±30° relative to the beam axis. $^{40}$Ca (≈8.2% enrichment) was a major contaminant in the $^{48}$CaCO$_3$ source material, and $^{40}$Ca$^{5+}$ has a nearly identical mass-to-charge ratio to the $^{48}$Ca$^{6+}$ ion, such that both ions were accelerated by the cyclotron. In the Rutherford scattering energy spectra, two peaks were observed and assigned to $^{40}$Ca and $^{48}$Ca, simplifying correction to the beam dose for the contaminant. Energy loss calculations were performed in LISE++ [22, 23] using the method of Ziegler [24, 25] as implemented in LISE++. In-flight EVR separation was based on magnetic rigidity in the achromatic section of MARS followed by a Wien filter. EVR charge state distributions were estimated according to Schiwietz and Grande [26]. The estimated magnetic rigidities ($B\rho$) and velocities ($v/c$) of the products during the experiments are shown in Table 2.

Implanting EVRs and their subsequent $\alpha$-decay were detected using a 16-strip position-sensitive Micron Semiconductor model X-1 Si focal plane detector. A four-peak alpha source with isotopes of $^{148}$Gd, $^{239}$Pu, $^{241}$Am, and $^{244}$Cm was used to obtain an external calibration. Internal calibrations (corrected for recoiling daughter energy) were obtained using the $\alpha$-decaying EVR products of the $^{48}$Ca + $^{165}$Ho, $^{45}$Sc + $^{118}$Sn, $^{50}$Ti + $^{106}$Pd, and $^{54}$Cr + $^{106}$Pd reactions (as appropriate). The energy resolution of the experimental set-up (full-width at half-maximum of a peak) was between 75–80 keV. In the $^{48}$Ca irradiations, data acquisition continued for 30 minutes after finishing an irradiation run to allow longer-lived implanted EVRs to decay.

In the experiments using $^{45}$Sc projectiles, a microchannel plate (MCP) detector was mounted upstream of the focal plane detector. Coincident signals in the MCP and focal plane detectors indicated the presence of an implanting ion. A focal plane signal anti-coincident with the MCP indicated a radioactive decay. The efficiency for identifying an implanting ion was >99%. In the $^{48}$Ca, $^{50}$Ti, and $^{54}$Cr experiments, the beam was pulsed, and beam ON and beam OFF spectra were recorded. The beam pulsed between equal time intervals by shifting the phase of one of the K500 dees by ≈10°, taking ≈0.5 ms to completely shut-off on each cycle. Beam ON spectra included all irradiation events (i.e., target and projectile-like fragments, scattered beam, etc.), while beam OFF spectra included only radioactive decays of EVRs implanted in the focal plane detector during the beam ON interval. A pulsing interval of 500 ms ON, 500 ms OFF was chosen, based on the half-lives of the reaction products of the $^{162}$Dy($^{48}$Ca, $xn$)$^{206-x}$Rn reaction; this resulted in an equal probability of observing α-decay events in both the ON and OFF windows. The same pulsing window was used for the $^{159}$Tb($^{50}$Ti, $xn$)$^{209-x}$Fr reaction. The interval was reduced to 50 ms ON, 50 ms OFF for the $^{162}$Dy($^{54}$Cr, $xn$)$^{212-x}$Th reaction due to shorter half-lives of the products. The energy spectrum recorded in the beam OFF period, having reduced background, allowed for accurate isotope identification and production rate determination.

**Table 2.** Kinematics of reactions and charge states studied in the present work. These data were used to determine settings for MARS. $v$ is the primary product's velocity, $B\rho$ is its magnetic rigidity, and $c$ is the speed of light.

| Reaction | Primary Product(s) | Charge State | $v/c$ (%) | $B\rho$ (T m) |
|---|---|---|---|---|
| $^{162}$Dy($^{48}$Ca, $xn$)$^{210-x}$Rn | $^{205,\,206}$Rn | 21+ | 1.90–2.10 | 0.58–0.64 |
| $^{159}$Tb($^{45}$Sc, $xn$)$^{204-x}$Rn | $^{200}$Rn | 21+ | 2.00–2.12 | 0.59–0.63 |
| $^{162}$Dy($^{45}$Sc, $xn$)$^{207-x}$Fr | $^{203}$Fr | 21+ | 2.01–2.12 | 0.61–0.64 |
| $^{159}$Tb($^{50}$Ti, $xn$)$^{209-x}$Fr | $^{205}$Fr | 21+ | 2.12–2.21 | 0.64–0.67 |
| $^{162}$Dy($^{54}$Cr, $xn$)$^{216-x}$Th | $^{212}$Th | 24+ | 2.34–2.39 | 0.64–0.66 |

## 3. Results

The excitation functions for the production of $4n$ reaction products are shown in Figure 1. (In the case of $^{48}$Ca + $^{162}$Dy, the data points indicate the total production of the $4n$ and $5n$ products because the two product nuclides emit alpha particles with similar energies that could not be distinguished with our equipment). In some cases, a clear peak above background could not be observed, so upper limit cross sections are indicated. These upper limits are calculated at the 84% confidence level using methods described in [27]. Error bars are shown at the $1\sigma$ level, also calculated according to methods described in [27].

In all cases, reactions with $Z_p > 20$ have cross sections that are smaller than those in the $^{162}$Dy($^{48}$Ca, $xn$)$^{210-x}$Rn ($x$ = 4,5) reactions by at least one order of magnitude. For example, although only upper limits were determined for the $^{162}$Dy($^{54}$Cr, $4n$)$^{212}$Th excitation function, the ratio of peak cross sections in these two reactions differ by a factor of $>7.1 \times 10^3$. Excluding the $^{48}$Ca + $^{162}$Dy reaction, the largest cross section measured in the current study is 0.24 mb for the $^{159}$Tb($^{50}$Ti, $4n$)$^{205}$Fr reaction at a laboratory-frame center-of-target projectile energy $E_{\text{cot}}$ of 220.8 MeV. This is a factor of 33 smaller than the peak cross section observed in the $^{48}$Ca + $^{162}$Dy reaction (although the latter reaction includes the sum of two exit channels as discussed above). These data could have significant implications for the production of SHEs as discussed below.

In some cases (especially the $^{45}$Sc-induced reactions), the production of $pxn$ products was substantially greater than the production of $xn$ products. This is likely due to the very neutron-deficient CN which are produced in these reactions. The neutron binding energies are very high and the proton binding energies are very low, so the emission of protons is probable even though proton emission is hindered by a high Coulomb barrier. Additionally, the emission of a high-energy (≈20

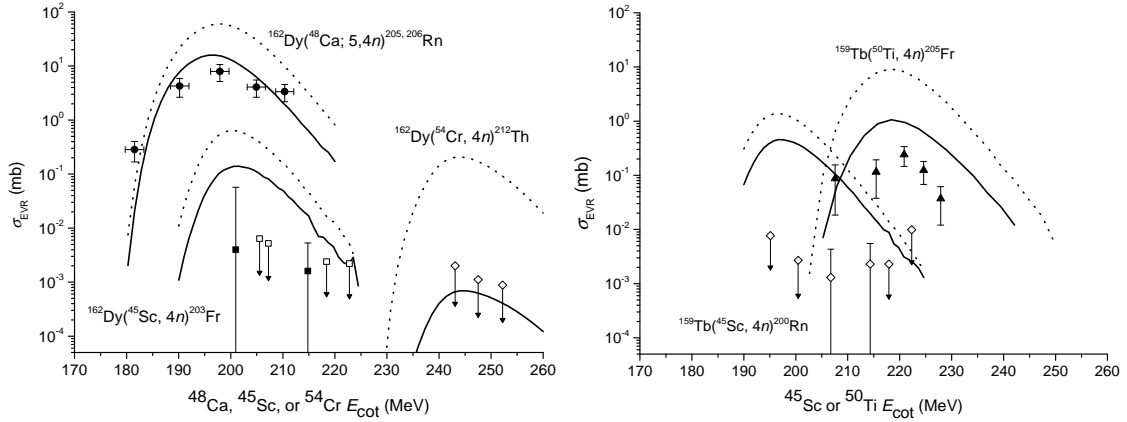

**Figure 1.** Measured 4n excitation functions for the reactions of $^{48}$Ca (circles), $^{45}$Sc (squares), $^{50}$Ti (triangles), and $^{54}$Cr (diamonds) with targets of $^{162}$Dy (left figure) and $^{159}$Tb (right figure). Horizontal error bars represent the range of energies covered in the target, and are only shown on the $^{48}$Ca + $^{162}$Dy reaction for clarity; energy losses in other targets are comparable. $E_{cot}$ is the laboratory-frame center-of-target projectile energy. Solid symbols indicate measured cross sections while open symbols indicate upper limits. Dotted lines are theoretical calculations without collective effects included, and solid lines show the same calculations with collective effects included. (See the main text for a discussion).

MeV) proton reduces the excitation substantially, increases the survival probability, and leads to a higher reaction cross section. A ($^{45}$Sc, $pxn$) reaction leads to the formation of a residue with the same atomic number as a ($^{48}$Ca, $xn$) reaction when the same target is irradiated. This fact, and the resulting substantial decrease in cross section, would seem to negate the justification for using a projectile with $Z_p > 20$. The analysis of the cross sections for $pxn$ reactions are ongoing as of the time of writing of this proceeding, and will be discussed in a future full publication.

## 4. Discussion

### 4.1. General theory of fusion-evaporation reactions

Typically, the cross section for a fusion-evaporation reaction $\sigma_{xn}$ is represented as the product of three terms, the capture cross section $\sigma_{cap}$, the probability of CN formation $P_{CN}$, and the probability of EVR survival $W_{sur}$:

$$\sigma_{xn} = \sigma_{cap} P_{CN} W_{sur}(E^*). \tag{1}$$

The survival probability can be estimated with transition state theory, and is commonly approximated using the probability of neutron emission from the CN:

$$W_{sur} = \prod_{i=1}^{x} (\Gamma_n / \Gamma_{tot})_i \approx \prod_{i=1}^{x} (\Gamma_n / \Gamma_f)_i, \tag{2}$$

where $\Gamma_n$ is the width for neutron emission from excited nucleus, $\Gamma_{tot}$ is the total width for decay by any mode, $\Gamma_f$ is the width for fission, and $x$ is the number of neutrons emitted. The approximation is justified since neutron emission and fission are the dominant decay modes, and $\Gamma_n \ll \Gamma_f \approx \Gamma_{tot}$.

Standard estimates of $\Gamma_n/\Gamma_f$ can be made using transition state theory, and lead to expressions such as [28]:

$$\frac{\Gamma_{\mathrm{n}}}{\Gamma_{\mathrm{f}}} = \frac{2m_{\mathrm{n}}gr_0^2 A^{2/3}}{\hbar^2} \frac{\int_0^{E^*-B_n} \varepsilon \rho_{\mathrm{n}}(E^* - B_n - \varepsilon)d\varepsilon}{\int_0^{E^*-B_f} \rho_{\mathrm{f}}(E^* - B_f - K)dK}, \qquad (3)$$

where $m_{\mathrm{n}}$ is the mass of the neutron, $g = 2$ is the spin degeneracy of the neutron, $r_0$ is the radius parameter, $A$ is the mass number of the emitting nucleus, $\varepsilon$ is the energy of the emitted neutron, $\rho_{\mathrm{n}}$ and $\rho_{\mathrm{f}}$ are the level densities of the system for neutron emission and fission, respectively, $B_{\mathrm{n}}$ is the binding energy of the neutron, $B_{\mathrm{f}}$ is the fission barrier, and $K$ is the kinetic energy of the fissioning system across the saddle point. In practice, a canonical expansion of the integrals in the numerator and denominator around $E^* - B_{\mathrm{n}}$ and $E^* - B_{\mathrm{f}}$, respectively, result in standard integrals which can be evaluated to give

$$\Gamma_{\mathrm{n}}/\Gamma_{\mathrm{f}} \propto \exp[-(B_{\mathrm{n}} - B_{\mathrm{f}})/T], \qquad (4)$$

where $T$ is the nuclear temperature. This Boltzmann-like result shows that the difference in neutron binding energy and fission barrier height should have a tremendous impact on the reaction cross section. This difference is discussed in further detail below.

The capture cross section is believed to be understood relatively accurately because data is available for comparison with theoretical predictions; calculations are believed to have an accuracy better than one order of magnitude for above-barrier reactions [29]. The transition state theory is based on a solid theoretical foundation and the approximations used above are believed to be reasonable. Thus, the largest uncertainty in the calculation of super heavy element formation cross sections is associated with $P_{\mathrm{CN}}$, estimates of which can be in error by as much as one order of magnitude [30]. The combined uncertainties of these factors are significant, such that predicted production cross sections for EVRs with $Z > 112$ are likely only accurate to two orders of magnitude [29].

*4.2. Role of $B_n - B_f$*

Equation (4) suggests that those CN which have the smallest values of $B_{\mathrm{n}} - B_{\mathrm{f}}$ will have the greatest chance of survival against fission. The fission barrier plays an especially important role that has been understood for many years. For example, Randrup *et al.* predicted in 1976 that even-even nuclei with $Z = 108$ should have spontaneous fission half-lives of $\approx 1$ µs [31]. These predictions were proven untrue with the discovery of the deformed subshell at $Z = 108$ and $N = 162$. (Figure 2 in [32] gives an indication of the strength of the shell correction energy in this subshell, which is up to 7 MeV). The evidence for this subshell includes the unusually long half-lives of nuclei such as $^{257}$Rf (4.7 s) [17] and $^{270}$Hs (22 s) [33], the low $Q_\alpha$ values for nuclei with $N = 162$ (see Figure 1 in [33]), and the relatively large cross sections for production of nuclei in that region in so-called "cold fusion" reactions with large values of $Z_pZ_t$. (See [34] for a review). The latter are especially important for establishing the importance of shell stabilization on reaction cross section. The existence of the deformed subshell provides an example of a "silver bullet" that has enabled experimenters to produce and study nuclei that, in principle, should be much more unstable than they actually are. In this case, stability is increased by the large shell correction energy, which helps maximize $B_{\mathrm{f}}$.

The next silver bullet which allowed experiments to study a new group of heavy elements was the use of $^{48}$Ca projectiles reacting with actinide targets. Intense beams of $^{48}$Ca became available at several laboratories in the 1990s, and these reactions ultimately led to the discovery of six new elements as described in the Introduction. This projectile has an unusually high neutron-to-proton ratio, and the resulting CN are more neutron-rich than those available with other projectiles. This results in lower $B_{\mathrm{n}}$, higher $B_{\mathrm{f}}$, and lower $B_{\mathrm{n}} - B_{\mathrm{f}}$, all of which increase $W_{\mathrm{sur}}$.

Table 3 shows the estimated values of $B_{\mathrm{n}} - B_{\mathrm{f}}$ for CN of the reactions of interest to this study, along with additional measures to characterize the reactions, including the mass asymmetry parameter $\eta$ (see the Table for a definition), the value of $Z_p Z_t$ in each case, and the effective quadrupole deformation of the CN $\beta_{2,\,CN}^{eff}$. Additional $B_n - B_f$ data for each step in the de-excitation cascade are shown in Figure 2. The values of $B_n - B_f$ for reactions studied in this work vary over a wide range (−7.05 to −0.84 MeV), ensuring that $W_{sur}$ will also vary over a wide range.

**Table 3.** Comparison of reactions of interest to this work. The subscripts "p" and "t" refer to the projectile and target, respectively. $\beta_{2,\,CN}^{eff}$ represents the effective quadrupole deformation of the compound nucleus, and $[B_n - B_f]_{CN}$ represents the difference in neutron binding energy ($B_n$) and fission barrier height ($B_f$) of the compound nucleus. $B_n$ data are taken from [17]. $B_f$ data represent the sum of a macroscopic component from [35] and a shell correction from [36]. $\beta_{2,\,CN}^{eff}$ is taken from [37] and assumes that the deformation in the excited state is the same as in the ground state.

| $Z_p$ | CN Reaction | $\eta = \dfrac{|A_p - A_t|}{A_p + A_t}$ | $Z_p Z_t$ | $\beta_{2,\,CN}^{eff}$ | $[B_n - B_f]_{CN}$ (MeV) |
|---|---|---|---|---|---|
| 20 | $^{48}$Ca + $^{159}$Tb → $^{207}$At$^*$ | 0.536 | 1300 | −0.035 | −7.20 |
|    | $^{48}$Ca + $^{162}$Dy → $^{210}$Rn$^*$ | 0.543 | 1320 | −0.026 | −7.05 |
|    | $^{48}$Ca + $^{165}$Ho → $^{213}$Fr$^*$ | 0.549 | 1340 | +0.008 | −6.81 |
| 21 | $^{45}$Sc + $^{159}$Tb → $^{204}$Rn$^*$ | 0.558 | 1365 | −0.087 | −1.37 |
|    | $^{45}$Sc + $^{162}$Dy → $^{207}$Fr$^*$ | 0.565 | 1386 | −0.104 | −1.59 |
|    | $^{45}$Sc + $^{165}$Ho → $^{210}$Ra$^*$ | 0.571 | 1407 | −0.053 | −1.49 |
| 22 | $^{50}$Ti + $^{159}$Tb → $^{209}$Ra$^*$ | 0.522 | 1430 | −0.044 | −2.75 |
|    | $^{50}$Ti + $^{162}$Dy → $^{212}$Ra$^*$ | 0.528 | 1452 | −0.035 | −2.68 |
|    | $^{50}$Ti + $^{165}$Ho → $^{215}$Ac$^*$ | 0.535 | 1474 | 0.000 | −2.70 |
| 23 | $^{51}$V + $^{159}$Tb → $^{210}$Ra$^*$ | 0.514 | 1495 | −0.053 | −1.49 |
|    | $^{51}$V + $^{162}$Dy → $^{213}$Ac$^*$ | 0.521 | 1518 | −0.044 | −1.17 |
|    | $^{51}$V + $^{165}$Ho → $^{216}$Th$^*$ | 0.527 | 1541 | +0.008 | −1.45 |
| 24 | $^{54}$Cr + $^{159}$Tb → $^{213}$Ac$^*$ | 0.493 | 1560 | −0.044 | −0.76 |
|    | $^{54}$Cr + $^{162}$Dy → $^{216}$Th$^*$ | 0.500 | 1584 | +0.008 | −0.84 |
|    | $^{54}$Cr + $^{165}$Ho → $^{219}$Pa$^*$ | 0.507 | 1608 | −0.008 | +1.73 |
| 25 | $^{55}$Mn + $^{159}$Tb → $^{214}$Th$^*$ | 0.486 | 1625 | −0.052 | +0.35 |
|    | $^{55}$Mn + $^{162}$Dy → $^{217}$Pa$^*$ | 0.493 | 1650 | +0.000 | +0.37 |
|    | $^{55}$Mn + $^{165}$Ho → $^{220}$U$^*$ | 0.500 | 1675 | +0.008 | +0.76 |

*4.3. Collective enhancements to level density*
In addition to $B_n - B_f$, collective effects can have a dramatic impact on the fusion-evaporation cross section. (See [38, 39] and references therein for a discussion). These enhancements to level density are caused by collective rotations (and to a lesser extent, vibrations) which can occur in excited nuclei.

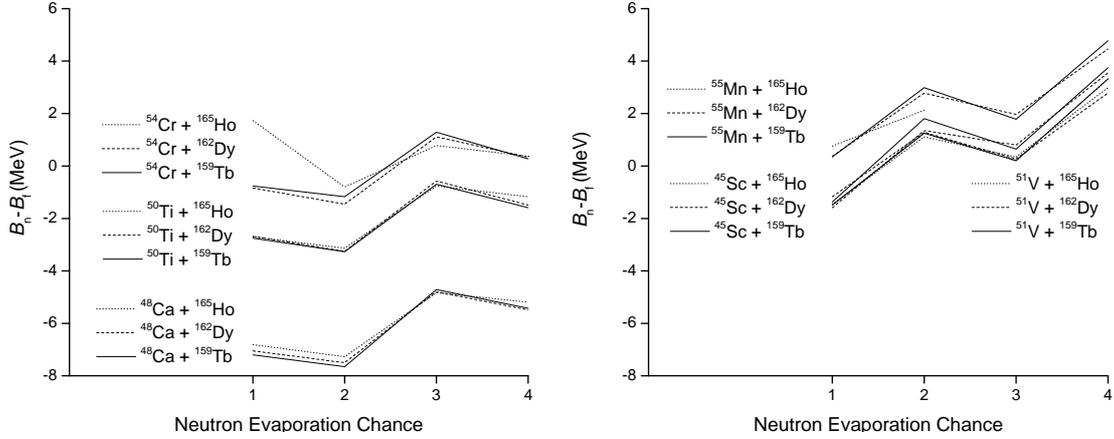

**Figure 2.** (left) $B_n - B_f$ for reactions of even-Z projectiles of interest to this work. (right) $B_n - B_f$ for reactions of odd-Z projectiles of interest to this work. The $^{45}$Sc- and $^{51}$V-induced reactions have virtually identical values of $B_n - B_f$. The $^{55}$Mn + $^{165}$Ho reaction only has two data points because the final two nuclei are unbound. The $^{45}$Sc + $^{165}$Ho reaction has the same CN as the $^{51}$V + $^{159}$Tb reaction. "Neutron Evaporation Chance" refers to first-chance neutron emission, second-chance neutron emission, etc., from the CN. The first chance data are also given in Table 3.

For example, a deformed excited nucleus will have a greater level density than a similar spherical nucleus with the same excitation energy. This occurs because the deformed nucleus can have rotational bands, but these are forbidden in spherical nuclei. During the initial stages of fission, a nucleus must increase its deformation, and the appearance of rotational bands increases its level density at the fission saddle. This increase in level density causes an increase in $\Gamma_f$, resulting in a decrease in survivability ($W_{sur}$). This effect is greatest for spherical nuclei, which can see a substantial increase in level density as the nucleus moves toward the saddle configuration. In contrast, a deformed nucleus would be less susceptible to this effect. Many spherical nuclei are located near closed shells, so the formation of such nuclei could be hindered relative to deformed nuclei.

The influence of collective effects can be seen in the present results. Figure 1 (left) shows calculations of the reaction cross section for the $^{162}$Dy($^{48}$Ca, $xn$)$^{210-x}$Rn ($x = 4, 5$) reactions both without (dotted line) and with (solid line) collective enhancements to level density included. These calculations were based on the formalism by Zagrebaev *et al.* [40, 41] using the proximity potential for the nucleus-nucleus interaction. In the $^{48}$Ca + $^{162}$Dy reaction, the calculation *with* collective enhancement provides substantially better agreement with the experimental data. The effect is even more pronounced in the case of the $^{162}$Dy($^{54}$Cr, $4n$)$^{212}$Th reaction, for which only upper limits were measured. When comparing the $^{48}$Ca, $^{54}$Cr + $^{162}$Dy reactions, $B_n - B_f$ for the CN changes by $\approx 6.2$ MeV and should result in a change in $W_{sur}$ by approximately two orders of magnitude. However, the difference in the experimental cross sections is a factor of $>7.1 \times 10^3$. The calculated effective quadrupole deformation $\beta_{2, CN}^{eff}$ of the $^{54}$Cr + $^{162}$Dy CN is very small (+0.008) [37]; if the same deformation were maintained during the de-excitation cascade then this nearly spherical nucleus would be expected to be especially susceptible to collective enhancements. This is indeed observed, as the difference in calculated cross section with and without collective enhancements is an additional factor of almost 300. The inclusion of collective enhancements results in predicted cross sections which are consistent with the experimental upper limit cross sections. These data show the important role played by collective effects in the survival of excited nuclei produced near closed shells.

The $^{45}$Sc-induced reactions have unusually small cross sections and are well below theoretical predictions, even with collective enhancements included. The cause of these very small cross sections is still under investigation.

*4.4. $P_{CN}$ estimates*

The data obtained in the current work allow for an estimation of $P_{CN}$ in some cases. If $\sigma_{cap}$ and $W_{sur}$ can be estimated, then the measured cross section $\sigma_{xn}$ can be used to calculate $P_{CN}$ according to Equation (1). Using the formalism by Zagrabaev *et al*. above to estimate $\sigma_{cap}$ and $W_{sur}$, $P_{CN}$ is estimated to be ≈0.5 for the $^{162}$Dy($^{48}$Ca, $xn$)$^{210-x}$Rn ($x = 4, 5$) reactions and ≈0.25 for the $^{159}$Tb($^{50}$Ti, $4n$)$^{205}$Fr reaction. In each case, the $P_{CN}$ estimate corresponds to the projectile energy at the peak of the excitation function. These estimates should be treated with significant caution as they likely have an error of at least one order of magnitude.

Although only upper limits were obtained for the excitation function of the $^{162}$Dy($^{54}$Cr, $4n$)$^{212}$Th reaction, similar reactions leading to the same EVR have been studied, including $^{176}$Hf($^{40}$Ar, $xn$)$^{216-x}$Th ($x = 4, 5$) [42], $^{152}$Sm($^{64}$Ni, $4n$)$^{212}$Th [43], $^{134}$Ba($^{82}$Se, $yn$)$^{216-y}$Th ($y = 4, 5$) [44], and $^{92}$Zr($^{124}$Sn, $zn$)$^{216-z}$Th ($z = 4, 5$) [45]. These reactions were found to have roughly constant maximum cross sections of ≈100 nb each, even though $\eta$ is very different in each case. Assuming that the cross section of the $^{162}$Dy($^{54}$Cr, $4n$)$^{212}$Th reaction is also ≈100 nb, $P_{CN} \approx 0.1$ can be estimated for this reaction. As above, this estimate likely has a very large error and should be treated with caution. Regardless, it does appear that there is some decrease in $P_{CN}$ with increasing $Z_p$, $Z_pZ_t$, and/or $\eta$.

## 5. Implications for the production of superheavy elements

The results of present study may have significant implications for the production of SHEs in current and future experiments. (Other authors have also discussed the production of heavy elements near subshell closures [38, 39, 42]). These results have shown that using projectiles with $Z_p > 20$ can result in a significant decrease in cross section relative to $^{48}$Ca-induced reactions. This appears to be due to a substantial increase in $B_n - B_f$ and a consequent reduction in $W_{sur}$ as shown in Equation (4). Unfortunately, the change from $^{48}$Ca to $^{45}$Sc, $^{50}$Ti, or $^{54}$Cr projectiles results in a CN which is more neutron-deficient, has larger values of $B_n$ during the de-excitation process, and has a greater probability of fissioning. Thus, $^{48}$Ca provides the benefits of a radioactive beam in the formation of SHEs due to its unusually high neutron-to-proton ratio, even though it is actually stable. In addition, CN which are more neutron-deficient are expected to be more fissile and have lower values of $B_f$. Siwek-Wilczyńska, Cap, and Wilczyński have estimated that a change in $B_f$ of ±1 MeV can change the evaporation residue cross section of a $3n$ reaction by a factor of 200 [18]. Any change in projectile away from $^{48}$Ca can have a number of negative consequences for the formation of SHEs. These results suggest that theoretical predictions [18, 46, 47] of sharp decreases in fusion-evaporation cross sections for reactions of $^{50}$Ti, $^{54}$Cr, $^{58}$Fe, or $^{64}$Ni with actinides compared to similar $^{48}$Ca-induced reactions are reasonable. These calculations typically predict maximum cross sections on the order of 10–100 fb.

A few words of caution may be in order here. The current study was focused on analog reactions, and the selection of these reactions may be shown in the future to have been unwise. Also, it is important that the reader not directly extrapolate the ratios of excitation functions studied in this work to SHEs. For example, although the peak cross sections of the $^{162}$Dy($^{48}$Ca, $xn$)$^{210-x}$Rn ($x = 4, 5$) reactions are greater than those of the $^{162}$Dy($^{54}$Cr, $4n$)$^{212}$Th reaction by a factor of $>7.1 \times 10^3$, this does not imply that a similar ratio will be observed between the $^{248}$Cm($^{48}$Ca, $4n$)$^{292}$Lv and reactions (if the excitation function of the latter were to be measured). In the case of the $^{162}$Dy reactions, the difference in $B_n - B_f$ is large (≈6.2 MeV), and such a large value is not expected in the $^{248}$Cm reactions. [Similar arguments can be made regarding the change from the $^{249}$Cf($^{48}$Ca, $4n$)$^{293}$118 reaction to the $^{249}$Cf($^{50}$Ti, $4n$)$^{295}$120 reaction]. Therefore, the change in $W_{sur}$ may not be as great as was observed in the current work. At the same time, estimates of $B_n$ for SHEs are uncertain, estimates of $B_f$ are even more uncertain, and estimates of $W_{sur}$ could be inaccurate. There may also be currently undiscovered influences which could make the production of SHEs more probable than is currently believed. Regardless, no such effect is anticipated.

In that sense, the "silver bullets" described in Section 4.2 may be exhausted. The first silver bullet, which was used to discover the first transactinide elements, was the deformed subshell near $Z = 108$ and $N = 162$. This subshell caused nearby nuclei to have increased $B_f$, and the fact that these nuclei are deformed may have helped to minimize $\Gamma_f$ due to a lack of collective enhancement. The second silver bullet was the use of neutron-rich $^{48}$Ca projectiles. This projectile is effectively a radioactive beam, although it is available at very high intensities. Unfortunately, the utility of this isotope for discovering new elements may be exhausted, since all available targets have been already been studied. [There is a small possibility that a suitable target of $^{253}$Es or $^{254}$Es could become available, but the discovery of element 119 in the $^{254}$Es($^{48}$Ca, $xn$)$^{302-x}$119 reaction would only further solidify the silver bullet nature of $^{48}$Ca]. It is not clear if another silver bullet is available which would allow for new elements to be discovered in the near future. The discovery of new elements with $Z > 118$ may require either a substantial improvement in accelerator technology (such as significantly increased primary beam intensities), or the use of a new reaction mechanism other than complete fusion-evaporation (such as massive multi-nucleon transfer reactions).

## 6. Conclusions

Excitation functions have been measured for the reactions of $^{48}$Ca, $^{45}$Sc, $^{50}$Ti, and $^{54}$Cr with $^{159}$Tb and $^{162}$Dy at the Cyclotron Institute at Texas A&M University using the Momentum Achromat Recoil Spectrometer. These reactions were chosen as analogs for the production of spherical superheavy nuclei because they require projectile energies above the Coulomb barrier, produce CN near spherical subshells, produce CN with similar excitation energies available for neutron emission, and produce CN with similar deformations to SHEs. The peak cross sections in the $^{45}$Sc-, $^{50}$Ti-, and $^{54}$Cr-induced reactions were substantially lower than the those for the $^{162}$Dy($^{48}$Ca, $xn$)$^{210-x}$Rn ($x = 4, 5$) reactions by at least one order of magnitude. Theoretical calculations were compared to the experimental data, and the inclusion of collective enhancements to level density resulted in better agreement in many cases. These results suggest that predictions of peak cross sections for the production of new elements with $Z > 118$ on the order of 10–100 fb are reasonable.

## 7. Acknowledgements

The authors wish to thank the Cyclotron Institute accelerator group and operations staff for delivering the beams of $^{48}$Ca, $^{45}$Sc, $^{50}$Ti, and $^{54}$Cr, and for assistance on numerous other issues. The authors wish to thank K Siwek-Wilczyńska and A V Karpov for many informative discussions. This work was supported by the United States Department of Energy under award numbers DE-FG02-93ER40773 and MUSC09-100, the Texas A&M University College of Science, the Welch Foundation under grant number A-1710, and the United States National Science Foundation under grant number PHY-1004780.

**References**
[1] Oganessian Y, 2007 *J. Phys. G* **34** R165-R242.
[2] Oganessian Y T, *et al.*, 2007 *Phys. Rev. C* **76** 011601.
[3] Oganessian Y T, *et al.*, 2000 *Phys. Rev. C* **62** 041604(R).
[4] Oganessian Y T, *et al.*, 2004 *Phys. Rev. C* **69** 021601-5.
[5] Oganessian Y T, *et al.*, 2002 *Eur. Phys. J. A* **15** 201-4.
[6] Oganessian Y T, *et al.*, 2010 *Phys. Rev. Lett.* **104** 142502.
[7] Oganessian Y T, *et al.*, 2006 *Phys. Rev. C* **74** 044602-9.
[8] Oganessian Y T, *et al.*, 2004 *Phys. Rev. C* **70** 064609.
[9] Oganessian Y T, *et al.*, 2009 *Phys. Rev. C* **79** 024603.

[10] Hofmann S, *et al.*, *Attempts for the Synthesis of New Elements at SHIP*, in GSI Scientific Report 2011 (2009); available at http://www-alt.gsi.de/informationen/wti/library/scientificreport2011/PAPERS/PHN-NUSTAR-SHE-01.pdf.
[11] Düllmann C E, *et al.*, *Upgrade of the Gas-filled Recoil Separator TASCA and First Search Experiment for the New Element 120 in the Reaction $^{50}Ti + ^{249}Cf$*, in GSI Scientific Report 2011 (2012); available at http://www-win.gsi.de/tasca/publications/annual_reports/annual_reports_2011/PHN-NUSTAR-SHE-02.pdf.
[12] Świątecki W J, Siwek-Wilczyńska K and Wilczyński J, 2005 *Phys. Rev. C* **71** 014602.
[13] Sobiczewski A, Gareev F A and Kalinkin B N, 1966 *Phys. Lett.* **22** 500.
[14] Meldner H, 1967 *Ark. Fys.* **36** 593.
[15] Rutz K, *et al.*, 1997 *Phys. Rev. C* **56** 238-43.
[16] Ćwiok S, *et al.*, 1996 *Nucl. Phys. A* **611** 211-46.
[17] *National Nuclear Data Center* (2012); available at http://www.nndc.bnl.gov.
[18] Siwek-Wilczyńska K, Cap T and Wilczyński J, 2010 *Intl. J. Mod. Phys. E* **19** 500-7.
[19] Tribble R E, Burch R H and Gagliardi C A, 1989 *Nucl. Instrum. Meth. A* **285** 441-6.
[20] Tribble R E, Gagliardi C A and Liu W, 1991 *Nucl. Instrum. Meth. B* **56/57** 956-9.
[21] Folden III C M, *et al.*, 2012 *Nucl. Instrum. Meth. A* **678** 1-7.
[22] Tarasov O B and Bazin D, 2004 *Nucl. Phys. A* **746** 411-4.
[23] Tarasov O B and Bazin D, *LISE++ Program*; available at http://www.nscl.msu.edu/lise.
[24] Ziegler J F, 2004 *Nucl. Instrum. Meth. B* **219-220** 1027-36.
[25] Ziegler J F, *Computer Code SRIM-2010*; available at http://www.srim.org.
[26] Schiwietz G and Grande P L, 2001 *Nucl. Instrum. Meth. B* **175-177** 125-31.
[27] Schmidt K-H, *et al.*, 1984 *Z. Phys. A* **316** 19-26.
[28] Vandenbosch R and Huizenga J R, 1973 *Nuclear Fission,* (New York: Academic Press) pp 227-33.
[29] Zagrebaev V I, *et al.*, 2001 *Phys. Rev. C* **65** 014607.
[30] Siwek-Wilczyńska K, *et al.*, 2008 *Intl. J. Mod. Phys. E* **17** 12-22.
[31] Randrup J, *et al.*, 1976 *Phys. Rev. C* **13** 229-39.
[32] Sobiczewski A and Pomorski K, 2007 *Prog. Part. Nucl. Phys.* **58** 292-349.
[33] Dvorak J, *et al.*, 2006 *Phys. Rev. Lett.* **97** 242501-4.
[34] Hofmann S and Münzenberg G, 2000 *Rev. Mod. Phys.* **72** 733-67.
[35] Sierk A J, 1986 *Phys. Rev. C* **33** 2039-53.
[36] Myers W D and Świątecki W J, 1996 *Nucl. Phys. A* **601** 141-67.
[37] Möller P, *et al.*, 1995 *Atom. Data Nucl. Data Tables* **59** 185-381.
[38] Armbruster P, 1999 *Rep. Prog. Phys.* **62** 465.
[39] Junghans A R, *et al.*, 1998 *Nucl. Phys. A* **629** 635-55.
[40] Zagrebaev V I, *et al.*, *Near-Barrier Fusion of Atomic Nuclei*; available at http://nrv.jinr.ru/nrv/.
[41] Zagrebaev V I, *et al.*, *Evaporation Residues*; available at http://nrv.jinr.ru/nrv/.
[42] Vermeulen D, *et al.*, 1984 *Z. Phys. A* **318** 157-69.
[43] Heredia J, *et al.*, 2010 *Eur. Phys. J. A* **46** 337-43.
[44] Ikezoe H, *et al.*, 2003 *Phys. Atom. Nucl.* **66** 1053-6.
[45] Sahm C C, *et al.*, 1985 *Nucl. Phys. A* **441** 316-43.
[46] Zagrebaev V and Greiner W, 2008 *Phys. Rev. C* **78** 034610.
[47] Liu Z H and Bao J-D, 2009 *Phys. Rev. C* **80** 054608.